\documentclass[a4paper,10pt,twoside]{cpc-hepnp}
\usepackage[latin9]{inputenc}
\usepackage[tbtags]{amsmath}
\usepackage{amssymb,bm,mathrsfs,bbm,amscd}
\usepackage{amssymb}
\usepackage{booktabs}
\usepackage{graphicx}
\usepackage{esint}
\usepackage{subfigure}
\usepackage{multirow}
\usepackage{lastpage}

\begin{document}

\fancyhead[c]{\small Chinese Physics C~~~Vol. xx, No. x (yyyy)
xxxxxx} \fancyfoot[C]{\small xxxxxx-\thepage}

\footnotetext[0]{Received dd mm yyyy}

\title{Conditioning of BPM pickup signals for operations of the Duke storage ring with a wide range of
single-bunch current\thanks{Supported by
US Department of Energy grant DE-FG02-97ER41033 and Fundamental Research Funds
for the Central Universities of China (WK2310000032)}}

\author{%
      XU Wei$^{1,2;1)}$\email{wxu@ustc.edu.cn}%
\quad LI Jing-Yi$^{1,2;2)}$\email{jingyili@ustc.edu.cn}%
\quad HUANG Sen-Lin$^{3}$%
\quad W.Z. Wu$^{2}$%
\quad H. Hao$^{2}$%
\quad P. Wang$^{2}$%
\quad Y.K. Wu$^{2}$%
}
\maketitle

\address{%
$^1$ National Synchrotron Radiation Laboratory (NSRL), University of Science and Technology of China, Hefei 230029, China\\
$^2$ Dept. of Physics, Duke University and FEL lab, Triangle Universities Nuclear Laboratory, Durham, NC 27708-0308, USA\\
$^3$ Institute of Heavy Ion Physics, Peking University, Beijing 100871, China
}

\footnotetext[0]{\hspace*{-3mm}\raisebox{0.3ex}{$\scriptstyle\copyright$}2013
Chinese Physical Society and the Institute of High Energy Physics
of the Chinese Academy of Sciences and the Institute
of Modern Physics of the Chinese Academy of Sciences and IOP Publishing Ltd}%

\begin{abstract}
The Duke storage ring is a dedicated driver for the storage ring
based oscillator free-electron lasers (FELs), and the High Intensity
Gamma-ray Source (HIGS). It is operated with a beam current ranging from
about 1 mA to 100 mA per bunch for various operations and accelerator physics studies.
High performance operations of the FEL and $\gamma$-ray source require a stable
electron beam orbit, which has been realized by the global orbit feedback
system. As a critical part of the orbit feedback system, the electron beam
position monitors (BPMs) are required to be able to precisely measure the
electron beam orbit in a wide range of the single-bunch current.
However, the high peak voltage of the BPM pickups associated with high single-bunch
current degrades the performance of the BPM electronics,
and can potentially damage the BPM electronics.
A signal conditioning method using low pass filters is developed
to reduce the peak voltage to protect the BPM electronics,
and to make the BPMs capable of working with a wide range of single-bunch current.
Simulations and electron beam based tests are performed.
The results show that the Duke storage ring BPM system is capable of providing
precise orbit measurements to ensure highly stable FEL and HIGS operations.
\end{abstract}

\begin{keyword}
Beam position monitor, Signal conditioning, Closed orbit feedback, Storage ring FEL
\end{keyword}

\begin{pacs}
29.20.db, 29.27.Bd
\end{pacs}

\section{Introduction}

The Duke storage ring is a dedicated driver for the storage ring based
free-electron lasers (FELs)~\cite{FELPRL}, and the High Intensity
Gamma-ray Source (HIGS)~\cite{HRWELLER2009}.
The high intense $\gamma$-ray beam is produced by colliding the high energy electron beam
with high power FEL optical beam inside the FEL cavity.
The electron beam orbit
in the storage ring not only impacts the beam injection and beam lifetime,
but also, especially in the FEL straight section which hosts the
optical klystron FEL, determines the quality of the FEL lasing
and impacts the flux
of the $\gamma$-ray beam~\cite{SUNTHESIS}. Maintaining good electron
beam orbit is essential for high performance operations of the Duke light
sources.

The electron beam orbit in the Duke storage ring is measured by the
beam position monitors (BPMs). This system was first brought
into operation in 1998~\cite{WANG1999}.
In the early days, the BPM measurement had
been found strongly depending on the single-bunch current.
Investigations showed that this effect was the result of overloading of the BPM electronics
modules due to the high peak voltage associated with a large single-bunch
current.
By employing cables with high loss in the GHz region, the
peak voltages were reduced, and the BPMs were able to provide reasonable
good orbit measurements for a single-bunch current ranging from 0.5 to 20
mA~\cite{WUPAC03}.

In 2006, a major upgrade of the light source facility was completed.
In the upgrade, a booster synchrotron was constructed as the full-energy
injector of the storage ring~\cite{SMPAC07}.
In addition, the north straight
section of the storage ring was overhauled for a new injection scheme
and for hosting a new high-order mode (HOM) damped RF cavity~\cite{WUPAC09,RFPAC05}.
In 2008, another key system of the storage ring,
the longitudinal bunch-by-bunch feedback (LFB) system,
was installed and commissioned for routine user operations~\cite{WWZPAC09,WWZNIMA11}.
Benefiting from all of these upgrades, the single-bunch current threshold
of the storage ring was significantly increased.
For example, a 95 mA electron beam was stored in the storage ring at 600 MeV in the
single-bunch mode with FEL lasing.
In the two-bunch HIGS operation mode,
the storage ring can be operated with a total beam current up
to 120 mA, or 60 mA in each bunch.
In these cases of high bunch-current operation,
the BPM electronics modules would have fully saturated without additional signal conditioning,
and were not able to perform meaningful orbit measurements~\cite{JLPAC07}.
Furthermore, the high single-bunch current produces
a very high peak voltage on the BPM pickup,
which could potentially damage the BPM electronics.

On the other hand, some operations need a low electron bunch current.
For example, some experiments need a small $\gamma$-ray beam energy spread,
which is usually realized by running a high-current bunch and a very low-current bunch
which does not produce FEL lasing.
Some machine studies also need a low beam current,
which sometimes is even lower than 1 mA.
In order to obtain precise measurement of the electron beam orbit in
a wide range of single-bunch current from about 1 mA to 100 mA,
the RF pickup signal must be
conditioned to increase the BPM's dynamic range.

Low pass filters (LPFs) have been widely used in BPM signal conditioning.
The purpose of this paper is to present a method using LPFs to increase the dynamic
range of the BPMs of the Duke storage ring to maintain high performance
of the storage ring and provide highly stable aiming of the $\gamma$-ray beam.
A preliminary study has been reported in a conference paper~\cite{JLPAC07}.
This paper first reports the measurements of two types of BPM pickups, button and stripline
pickups which are used in the Duke storage ring.
Then, it describes the details of the pickup signal conditioning.
The dependency of the BPM readings on the bunch current is measured and
reported in the later part of this paper.

\section{A revisit of BPM pickup signals}

Let us consider a relativistic electron bunch with a longitudinal
Gaussian distribution of its charges circulating in a storage ring.
The beam current is periodic and can be written in the time and frequency
domains as~\cite{WWZNIMA11},
\begin{eqnarray}
I_{\textrm{b}}\left(t\right) & = & \frac{I_{0}T_{0}}{\sqrt{2\pi}\sigma_{\tau}}
                       \sum\limits _{n=-\infty}^{\infty}e^{\frac{-\left(t-nT_{0}\right)^{2}}{2\sigma_{\tau}^{2}}},\nonumber \\
I_{\textrm{b}}\left(\omega\right) & = & I_{0}\sum\limits _{m=-\infty}^{\infty}e^{\frac{-\left(\omega-m\omega_{0}\right)^{2}}{2\sigma_{\omega}^{2}}},\label{eq:I}
\end{eqnarray}
respectively, where $I_{0}$ is the averaged beam current, $\sigma_{\tau}$ is the
RMS bunch length in the time domain,  $\sigma_{\omega}=1/\sigma_{\tau}$ is the RMS
spectral width, $T_{0}$ is the
revolution time of the electrons, $\omega_{0}=2\pi/T_{0}$ is the
angular revolution frequency, $n$ is the turn number, and $m$ is
the harmonic number of the revolution frequency. Eq.~(\ref{eq:I})
indicates that an electron beam with a short bunch length has a very
wide spectrum in the frequency domain.
As an example,
the natural RMS bunch length of an electron beam
with energy of $E_{\textrm{b}}$ = 1 GeV and RF voltage of $V_{\textrm{RF}}$ = 850
kV in the Duke storage ring is 2.0 cm~\cite{WUTHESIS},
its spectrum in the frequency domain
is comprised of a rich set of frequency responses from DC to GHz region.

The electron beam can be measured by BPM electrode pickups with a proper grounding.
Interacting with the electromagnetic
field associated with the electron beam, the electrode pickups can be used to
measure the electron beam current and position. Following the reports presented
by G.~R. Lambertson and D.~A. Goldberg on electromagnetic detectors
~\cite{GRLambertson1989,DAGoldberg1992},
we revisit the signals from two types of BPM pickups, button and stripline pickups which are used
in the Duke storage ring.

The measured pickup signal can be described by
\begin{equation}
V=Z_{\textrm{T}}I_{\textrm{b}}\textrm{,}\label{equ-bpmsignal}
\end{equation}
where $Z_{\textrm{T}}$ is the transfer impedance of the pickup.
A button pickup is an isolated
plate on the wall of the vacuum chamber. Compared to the electron
bunch length, the size of the plate is usually small.
It has a dominant capacitive
coupling and typically modeled with an equivalent RC circuit. When
an electron bunch passes the plate, the capacitor experiences charging
and discharging processes, the voltage across R is given by~\cite{GRLambertson1989,DAGoldberg1992},
\begin{equation}
V\left(\omega\right)=jgl\frac{\omega}{\beta c}\frac{R}{1+j\omega RC}I_{\textrm{b}}\left(\omega\right),\label{equ-button}
\end{equation}
and the transfer impedance of a button pickup is ~\cite{GRLambertson1989,DAGoldberg1992},
\begin{equation}
Z_{\textrm{T}}^{\text{button}}(\omega)=jgl\frac{\omega}{\beta c}\frac{R}{1+j\omega RC},\label{equ-button}
\end{equation}
where $R$ is the resistance of the circuit, $C$ is the capacitance
of the button pickup, $l$ is the effective length of the button,
and $g$ represents the ratio of the image charge on the plate to the
total image charge and is sensitive to the beam position.

A stripline pickup is a long plate with one end grounded on the wall
of the vacuum chamber, and the other end connected to the processing
electronics via an output line. The grounded end is usually in the
downstream to the electron beam motion. In this case, when an electron
bunch arrives the upstream end of the stripline, a signal is coupled
into the electrode and divided into two parts --- one part moves toward
the output line and the other part moves toward the downstream end.
When the electron bunch leaves the downstream end,
a signal with an opposite polarity is generated.
After canceling the signal
from the upstream, the signal moves toward the output line. Assuming
the electrons and the excited signals both move with the speed of
light $c$, there are two pulses moving through the output line separated
by $\Delta t=2l/c$, where $l$ is the effective length of the stripline.
Also assuming the electrode characteristic impedance and the output line
impedance are $Z_{L}$, the signal can be written as~\cite{GRLambertson1989,DAGoldberg1992},

\begin{equation}
V(t)=\frac{1}{2}gZ_{\textrm{L}}\left(I_{\textrm{b}}\left(t\right)-I_{\textrm{b}}\left(t+\frac{2l}{c}\right)\right),\label{equ-stpl_i}
\end{equation}
By Fourier transform, we can get the stripline's transfer impedance~\cite{GRLambertson1989,DAGoldberg1992},
\begin{equation}
Z_{\textrm{T}}^\textrm{stripline}=gZ_{\textrm{L}}e^{j\left(\pi/2-kl\right)}\textrm{sin}\left(kl\right),\label{equ-stripline}
\end{equation}
where $k=\omega/c$ is the wave number of the signal. When $kl=\pi/2$,
i.e. $l=\lambda/4$, $Z_{\textrm{T}}^\textrm{stripline}$ is a real number, and its
amplitude reaches maximum. For this reason, this type of stripline
BPM is sometimes called a quarter-wave loop device.

Equations (\ref{equ-bpmsignal}), (\ref{equ-button}) and (\ref{equ-stripline})
indicate that the spectrum of a BPM pickup signal is determined by
both the spectrum of the electron beam and transfer impedance of the
pickup electrode.

\section{The Duke Storage Ring BPM System}

The racetrack-shaped Duke storage ring is comprised of two arcs, the
east arc (EARC) and west arc (WARC), and two 34-meter long straight sections, the
north straight section (NSS) and south straight section (SSS). The
storage ring driven oscillator FEL is located in the SSS.
The NSS hosts the injection kickers, RF cavity, LFB
kicker, and other beam measurement systems.
The arc lattice is comprised
of ten FODO cells, including eight regular cells and one modified
cells at each end~\cite{WUTHESIS}.
Due to the space limitation, the sextupole magnets
were removed from the original design in the arc, and the sextupole
magnetic fields were realized by shifting the magnetic centers of the
quadrupole magnets horizontally inward by about 2.5 mm~\cite{VLPAC95}.
Consequently, this arrangement makes the electron beam have a large horizontal
offset at the locations of the arc BPM pickups, causing them to have a significant
nonlinear response to the beam orbit offset.

Thirty-three BPMs are currently used to measure the electron beam
orbit around the storage ring. Three types of electrodes, 30.4 mm long striplines (named LSL),
21.6 mm long short striplines (named SSL) and buttons with 5.0 mm diameter,
are used as BPM pickups. They are distributed around the storage ring as shown
in Table~\ref{pickup_dist}.
\begin{table}[htb!]
\begin{center}
\caption{The distribution of the 33 BPM pickups in the Duke storage ring.}
\vspace*{3mm}
\begin{tabular}{lcccc}
\toprule
&East Arc&West Arc&NSS&SSS\\\hline
\# of LSL    &0&0&6&0\\
\# of SSL    &9&9&0&2\\
\# of Button &0&0&1&6\\
\bottomrule
\end{tabular}
\label{pickup_dist}
\end{center}
\end{table}

Besides being used in the global orbit feedback, some BPMs in the
SSS are also used to allow local adjustments of
the electron beam orbit to optimize the FEL lasing
and $\gamma$-ray beam production of the HIGS.

The Bergoz multiplexer BPM modules are used to process the pickup
signals~\cite{Bergoz-spec}.
An on-board 1 GHz low pass filter (LPF) is used for each
of the four RF inputs to protect the module from being damaged by
very short, high voltage pulses. After the LPFs, the signals are time-multiplexed
into a superheterodyne receiver, and then fed through a band-pass
filter (BPF), which defines the operation range of the module.
Only those signal harmonics in the operation range are used for processing
the orbit information~\cite{Bergoz-spec}. The operation range of
the Duke storage ring BPM modules is 178.55 $\pm$ 20 MHz, which is
close to the frequency of the RF system.

A VME-based 64-channel 16-bit ADC board is used to digitalize the
orbit signal from the output of BPM modules. The orbit measurement
system is a part of the control system, which is based upon the Experimental
Physics and Industrial Control System (EPICS)~\cite{DFELCONTROL}.
The orbit data can be accessed via EPICS channel access (CA) at a
rate of 10 Hz or higher.

Low loss Heliax cables were originally used for connecting the BPM
pickups and electronics modules. It was found that the BPM modules
were highly overloaded with this arrangement.
RG223-U cables with higher loss in the GHz region were then used instead.
This enables the BPMs to provide reasonably good orbit measurements for a single-bunch current
ranging from 0.5 to 20 mA.
For a single-bunch current higher than
20 mA, most BPMs cannot provide precise measurements, and the pickup
signals need to be further conditioned.

\section{BPM pickup signal conditioning}

In order to learn about the properties of the pickup signals,
the time domain and frequency domain signals of the pickups are studied in this section.

\subsection{The time domain signals}

\begin{figure}[!hbt]
\centering \subfigure[]{ \includegraphics[clip,width=0.45\textwidth]{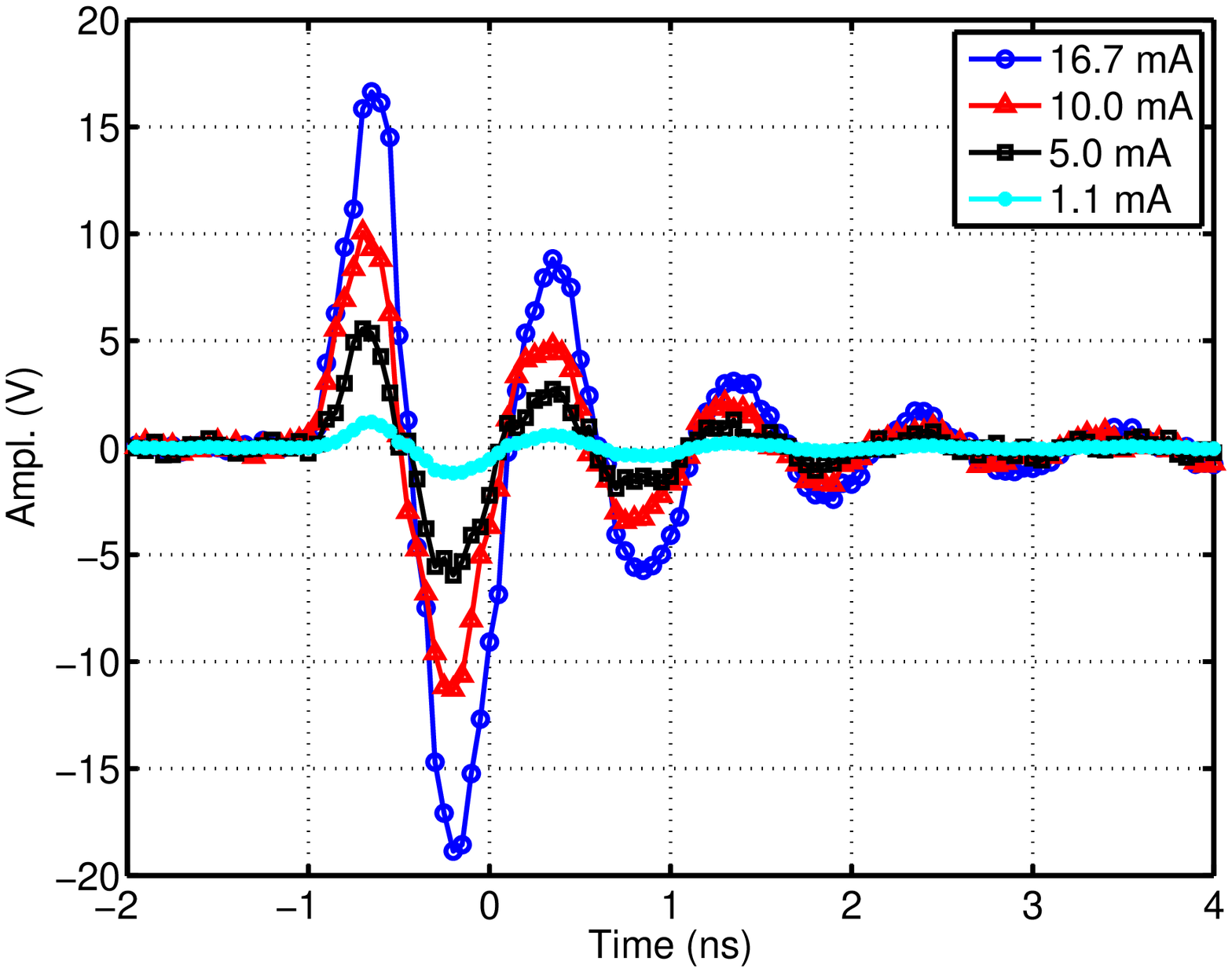}
\label{pv-1} }
\subfigure[]{ \includegraphics[clip,width=0.44\textwidth]{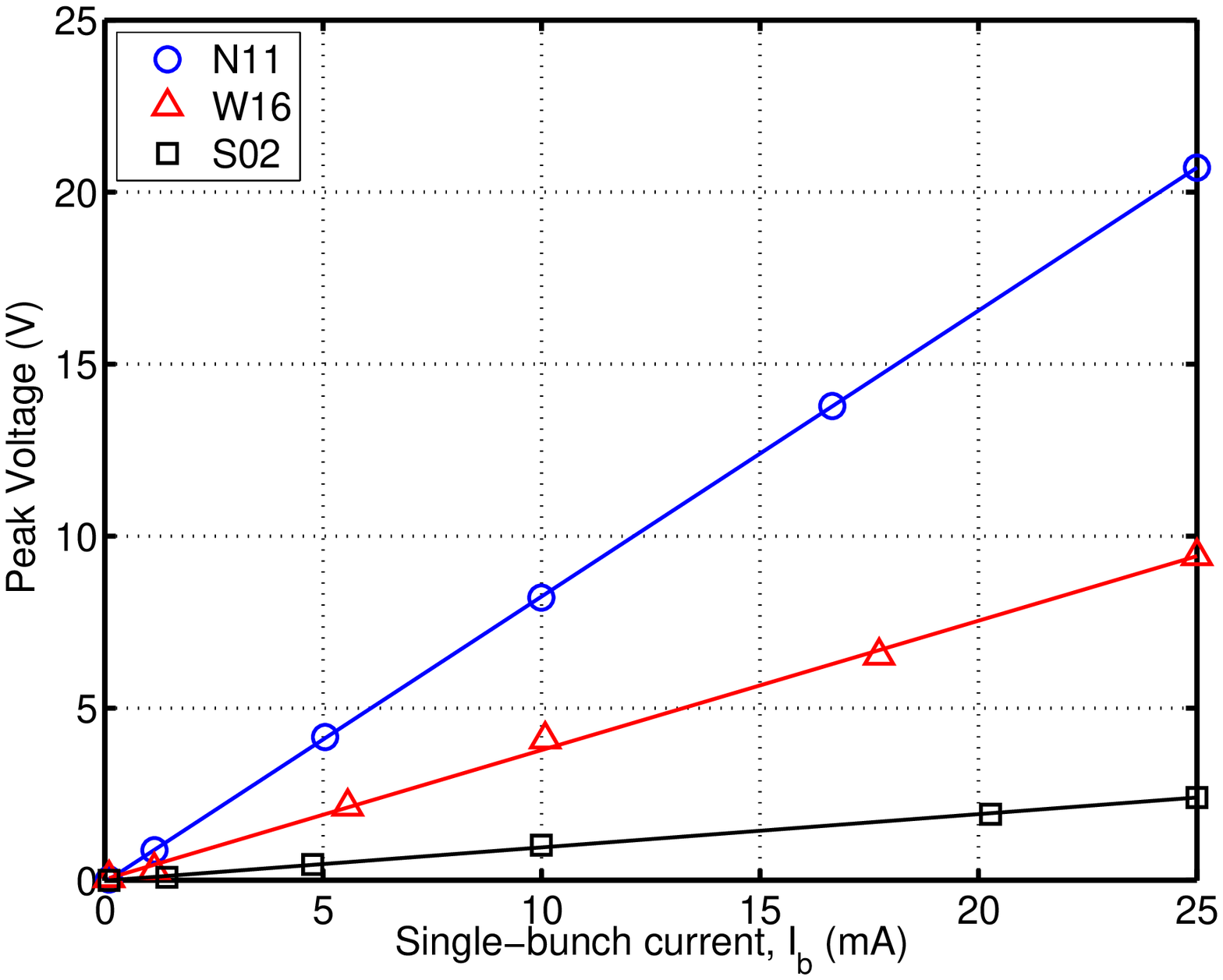}
\label{pv-2} }
\caption{(color online) Time domain BPM pickup signals. (a) Time domain signals
of an LSL electrode measured at different single-bunch currents. (b) Peak voltages
of the BPM pickup signals. The circles, stars and squares are peak
voltages obtained from the measured signals after applying a 1 GHz LPF.
The lines show the linear fits.
Measurements for all three types of BPMs (LSL, SSL, and button) are shown with
three BPMs from different sectors of the storage ring (N11, W16, and S01, respectively.)}
\label{bpm-peakVoltage}
\end{figure}

The beam signal from the BPM pickup is comprised of a series of very
short pules in the time domain, and thus has a very wide spectrum
in the frequency domain. In order to measure enough details of the
signal in the time domain, a digital oscilloscope, Tektronics TD7404
digital oscilloscope with a 4 GHz bandwidth and an up to 20 GHz sampling
rate, is used to measure the BPM pickup signals. A 60-feet long RG223-U
cable is used to connect the BPM pickup and the oscilloscope. To protect
the oscilloscope from high peak voltage, proper broadband attenuators
are used to bring the signal to a reasonable level. The signals from
each of the three types of the pickups are measured as a function
of single-bunch beam current.
As an example, a set of LSL pickup signals measured as a function of
the single-bunch beam current are shown in Fig.~\ref{pv-1}.
The ringing pattern of the signal is caused by the RF filtering the cable.
The peak voltages of BPMs from different sectors of the Duke storage ring,
N11, W16 and S02 which are LSL, SSL and button BPMs respectively,
are shown as a function of the single-bunch current in Fig.~\ref{pv-2}.

According to the manufacture's specifications,
in order to make the BPM
module work properly and avoid the electronics damage from high
peak voltages, the signal level after the 1 GHz on-board LPF of the
Bergoz BPM electronics module should be lower than 5 volt~\cite{WANGPRIVATE}.
To estimate the peak voltage after the on-board LPF, the measured
time domain signal is processed using a 1 GHz digital LPF. The results
show that the peak voltage after the digital LPF would exceed 5 volts around
6 mA for LSL BPMs and 13 mA for SSL BPMs. Assuming the peak voltage
is proportional to the single-bunch current, by extrapolating the data
of the button pickup signal, the peak voltage of button BPMs reaches 5 volts
around 52 mA, see Fig.~\ref{pv-2}. This means in order for the BPM
system to work at a single-bunch current up to 100 mA, all BPM pick
signals must be properly conditioned to reduce their peak voltages.

\subsection{Spectrum power distribution}

\begin{figure}[hbt!]
\centering
\subfigure[]{ \includegraphics[clip,width=0.45\textwidth]{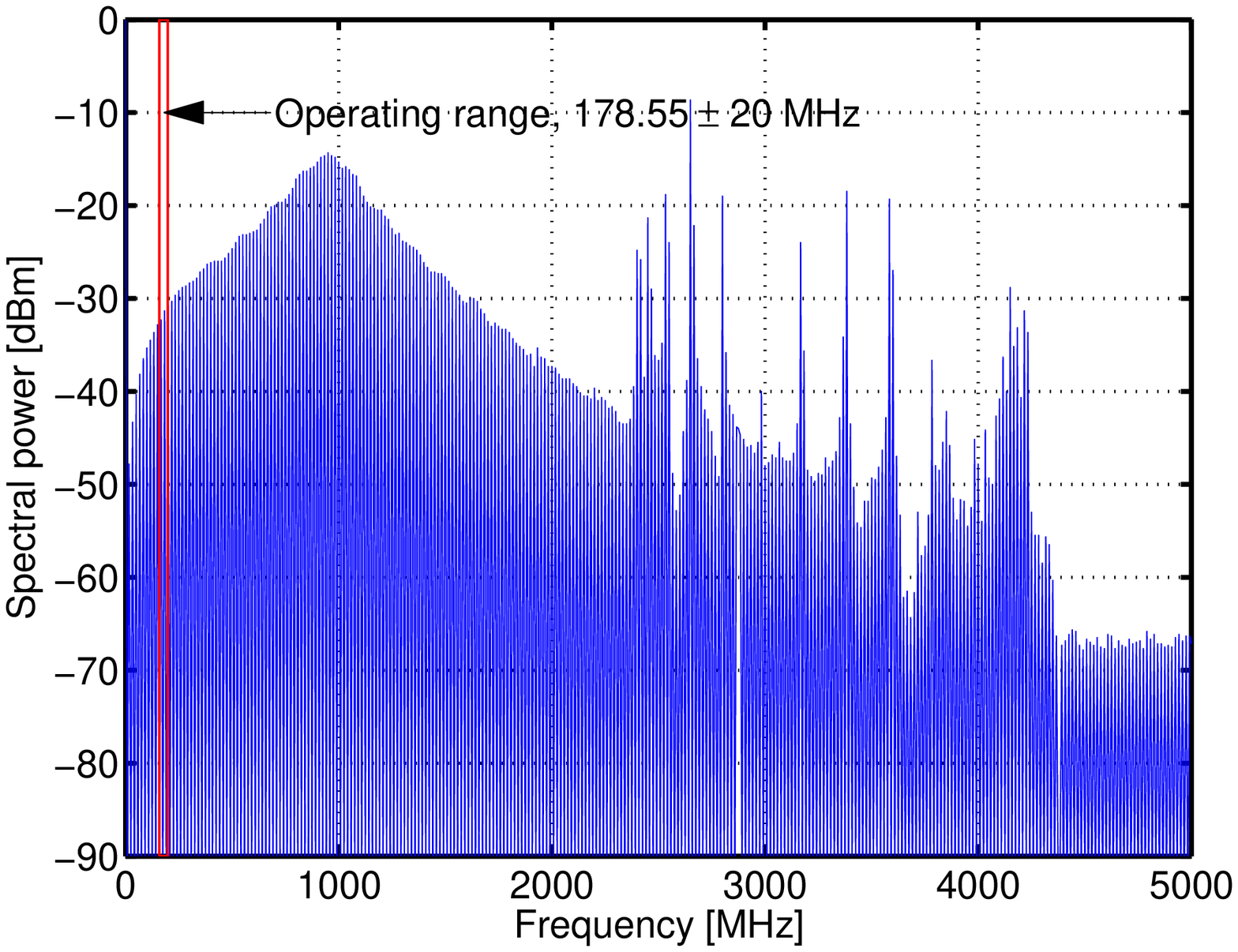}
\label{lsl-spec} }
\subfigure[]{ \includegraphics[clip,width=0.45\textwidth]{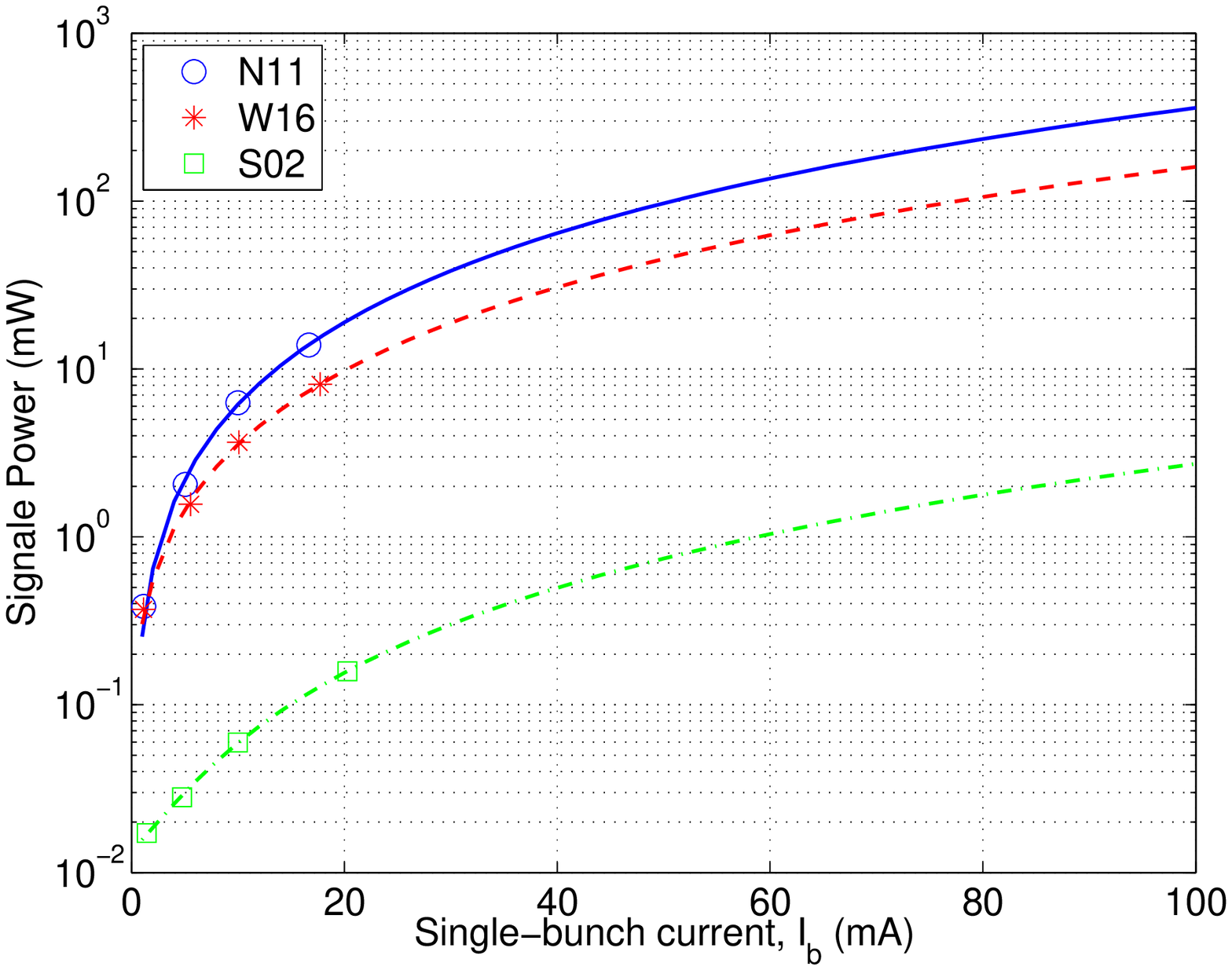}
\label{pwr-Ib} }
\caption{(color online) BPM pickup signals in the frequency domain. (a) A spectrum
of an LSL pickup signal. This spectrum is measured after a 60-feet
long RG223-U cable. (b) The total spectrum power as a function of the
single-bunch current. }
\label{bpm-spec}
\end{figure}

As discussed in the previous sections, the power of the pickup signal
is distributed in a very wide range in the frequency domain. The power
spectrum is determined by the characteristic impedance of the pickup
for a given beam signal.
To understand the spectral power distribution, the spectra
of the pickup signals are measured using an HP 8563E spectrum analyzer.
Figure~\ref{lsl-spec} shows a measured spectrum of the signal from
one of the LSL pickups. This figure indicates that only a very small
portion of the signal power is in the BPM's frequency operation range,
178 $\pm$ 20 MHz. In Fig.~\ref{pwr-Ib}, the measured and projected
total spectral power of all three types of pickups are plotted
as a function of the single-bunch current. The projected total powers at
100 mA of these pickups are listed in Table~\ref{bpm-t_pwr}.
Table~\ref{bpm-t_pwr} also lists the ratios of the power in
the BPM's operation range to the total power for a 20 mA single-bunch
beam.

\begin{table}[htb!]
\begin{center}
\caption{ The spectrum power of the pickup signals.
$P_{100}$ is the projected total power of a 100 mA
single-bunch current.
$P_{20}$ and $P_{20}'$ are the measured total power and the power in
the BPM's operation range, 178 $\pm$ 20 MHz, of a 20 mA single-bunch current,
respectively.}
\vspace*{3mm}
\begin{tabular}{lccc}
\toprule
 &LSL&SSL&Button\\\hline
$P_{100}$ [mW]&$\sim$360&$\sim$160&$\sim$3\\
$P_{20}'/P_{20}$&0.2\%&0.2\%&0.8\%\\
\bottomrule
\end{tabular}
\label{bpm-t_pwr}
\end{center}
\end{table}

Since only those revolution harmonics in the frequency operation range are actually
used for beam position processing, removing other harmonics by
a proper band pass filter (BPF) would have no significant adverse impact on the beam position
measurement. Because the operation frequency of the Duke storage ring
BPM is relatively low, an LPF can also do the job as a BPF.
After removing unused harmonics in the high frequency region, the
signal pulse length in the time domain is increased, the total power
of the signal is reduced, and hence the peak voltage is reduced.

\subsection{Simulation on the signal conditioning}
To get better understanding of the pickup signal conditioning using LPFs,
simulations using measured time domain signals and LPFs' insertion losses are performed.

The Duke storage ring has been operated with a variety of modes and bunch patterns.
Its single-bunch current varies by two orders of magnitude
to meet the requirements of various user programs.

\begin{figure}[htb!]
\centering
\subfigure[]{ \includegraphics[clip,width=0.45\textwidth]{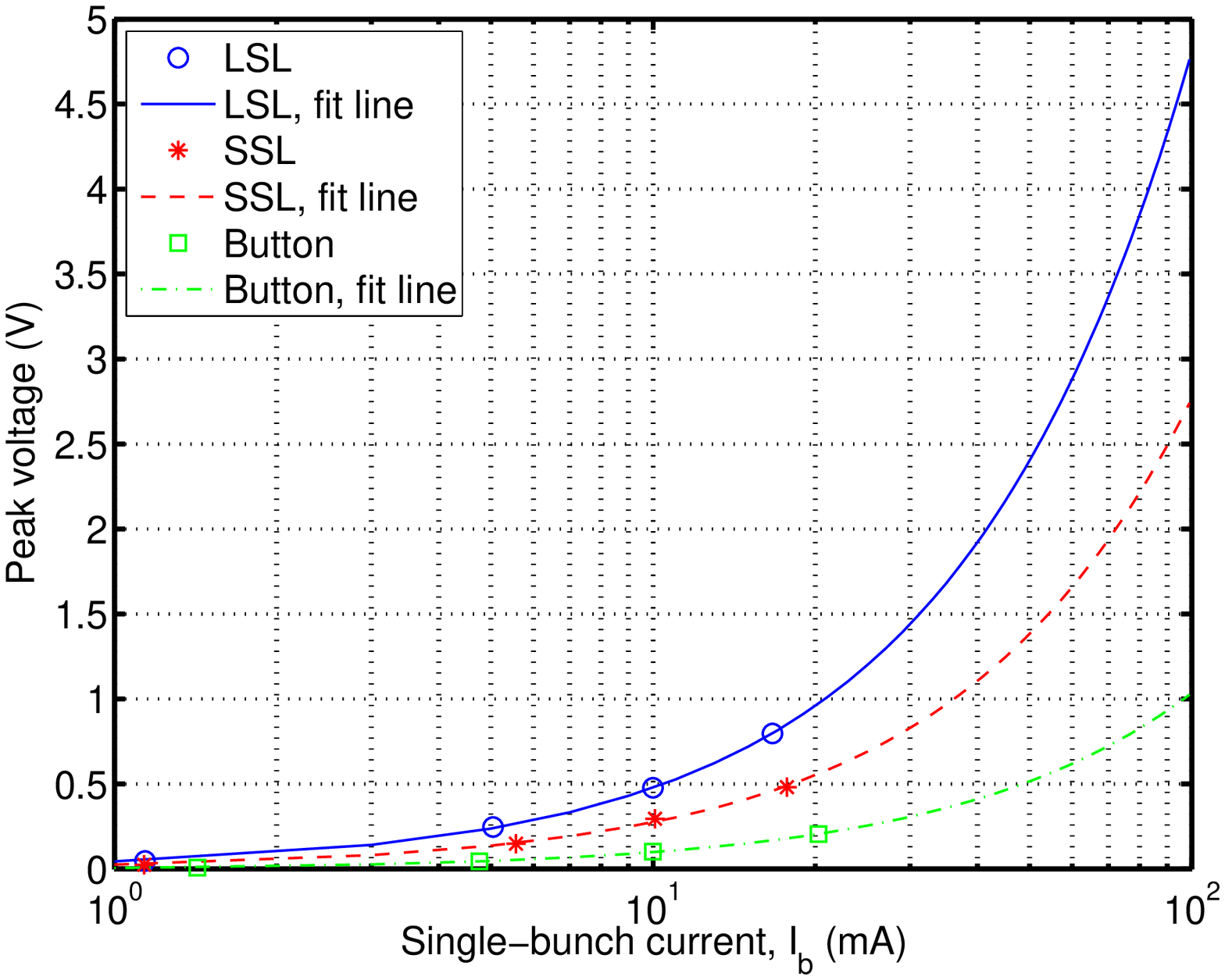}
\label{bpm-vp_vlf160} }
\subfigure[]{ \includegraphics[clip,width=0.45\textwidth]{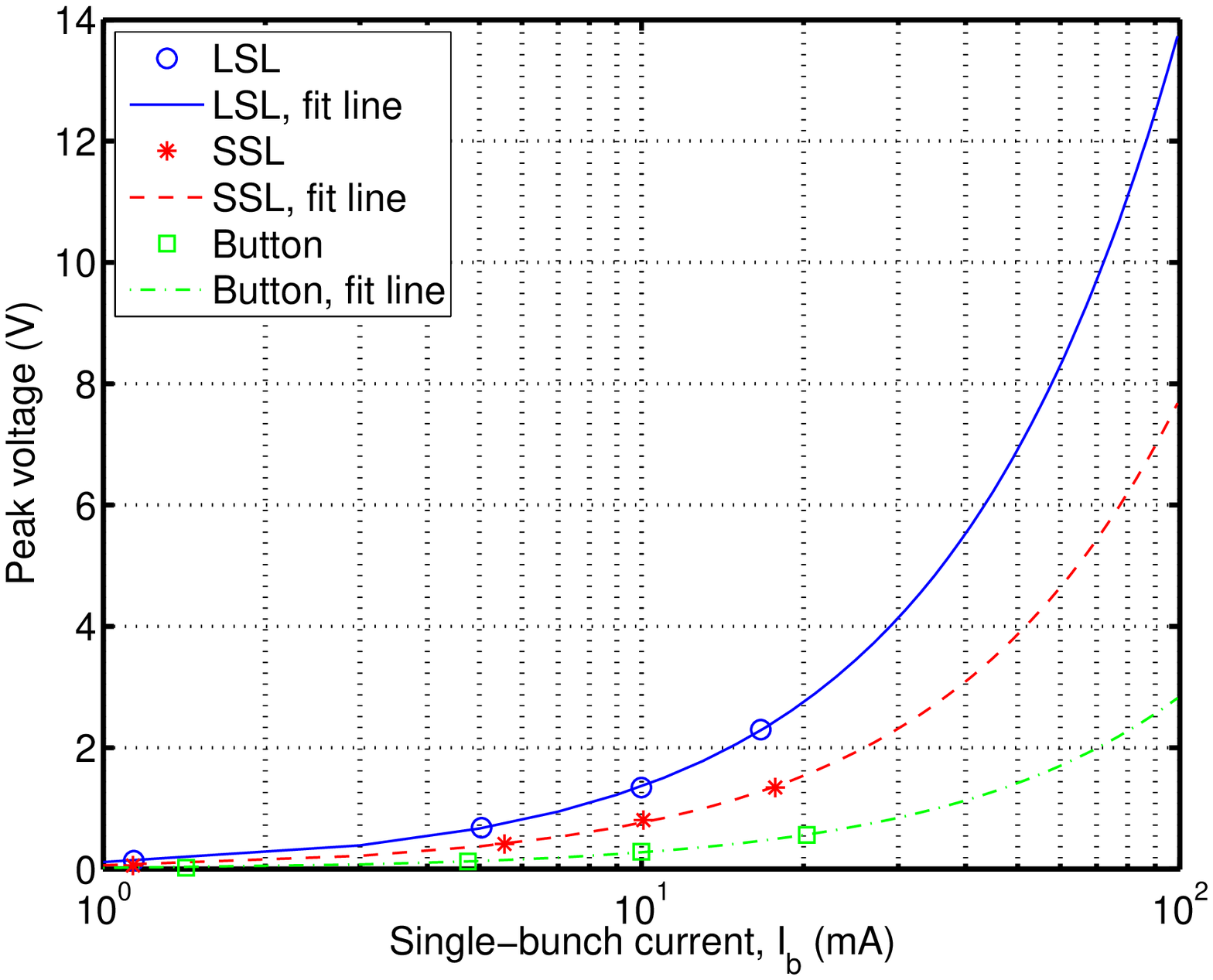}
\label{bpm-vp_vlf320} }
\caption{(color online) BPM pickup signal conditioning simulations using the measured
time domain signal and LPFs' specifications. (a) simulation results using
VLF-160 LPFs; (b) simulation results using VLF-320 LPFs.
The circle, star and square symbols are the peak voltage obtained from the measured
time domain signal after a 1 GHz digital LPF  for the LSL, SSL and button pickups,
respectively.
The solid, dashed and dot-dashed lines are their fit lines.
}
\label{bpm-vp_sim}
\end{figure}

To work with a high beam current, the filters used for the signal conditioning must have
a high input power threshold.
Low insertion losses around 178 MHz is essential to enable the BPM electronics
work with a low beam current.
With careful researches on the commercially available filters, two type of
Mini-circuits LPFs, VLF-160 and VLF-320, are found best meet these
requirements~\cite{VLF160}\cite{VLF320}.
Their input power thresholds are higher than 3 W.
The cut-off frequency (3 dB) of VLF-160 and VLF-320 LPFs are around 236 MHz and 460
MHz, respectively. Both of them have a small insertion
loss around 178 MHz, the operation frequency of the BPM electronics.

Using the manufacture's specifications on insertion loss and the
measured time domain signals, simulations are performed to evaluate
the peak voltage reduction. In these simulations, the measured time
domain data are transformed to the frequency domain using the fast Fourier
transform (FFT) to obtain the power spectrum.
After applying  the frequency filtering and power loss characteristics of
the LPFs to the power spectrum,
the frequency domain signal is transformed back to the time domain using the inverse FFT.
The peak voltage is calculated using the processed time domain data.
Figure~\ref{bpm-vp_sim} shows the calculated peak voltage as a function
of the single-bunch current for all three types of BPM pickups. By extrapolating
these curves, the peak voltages associated with a 100 mA single-bunch current
are calculated and tabulated in Table~\ref{tbl-pv_sim}.

\begin{table}[htb!]
\centering
\caption{Projected peak voltages of the pickup signals associated with a 100 mA single-bunch
current after conditioning with LPFs.}
\begin{tabular}{lccc}
\toprule
 & LSL & SSL & Button\tabularnewline
\hline
Using VLF-160 LPF & 4.8 V & 2.8 V & 1.0 V\tabularnewline
Using VLF-320 LPF & 13.9 V & 7.8 V & 2.8 V\tabularnewline
\bottomrule
\end{tabular}
\label{tbl-pv_sim}
\end{table}

These values are computed based upon the assumption that the electron
bunch length does not change with the single-bunch current. In practice,
the bunch length increases with the single-bunch current due to microwave
instability induced bunch lengthening.
However, this effect has no significant impact on the spectral power at low frequencies,
therefore, was neglected in the simulation studies.
Results in Table~\ref{tbl-pv_sim}
and Fig.~\ref{bpm-vp_sim} indicate that VLF-160 LPFs can be used to effectively
reduce the peak voltage at a high single-bunch current for all types
of pickups, while VLF-320 LPFs are only useful for the button pickup.

\subsection{Beam based test on signal conditioning}

The signal conditioning method discussed in the previous sections
\begin{figure*}[!hbt]
\centering
\subfigure[]{ \includegraphics[width=0.432\textwidth]{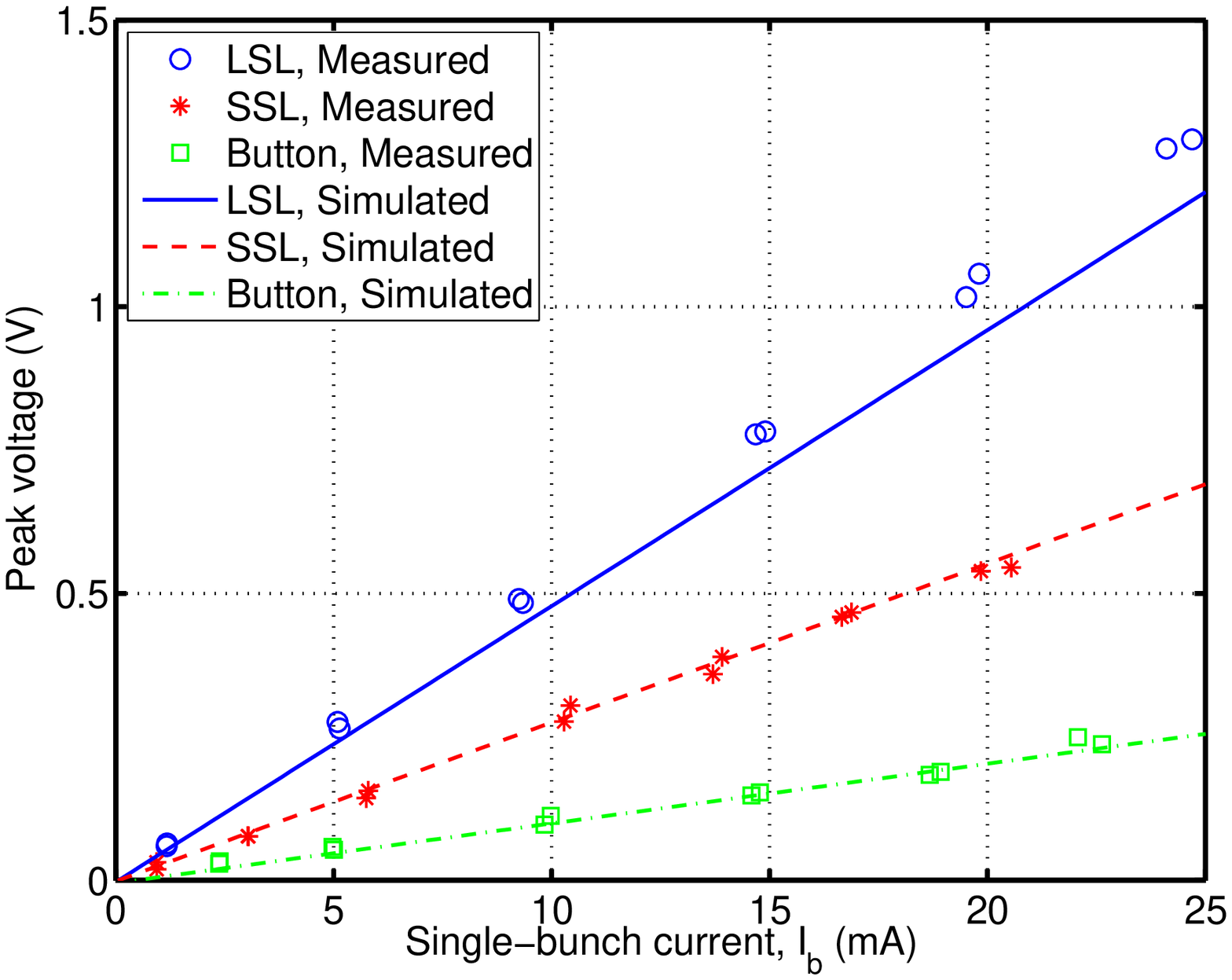}
\label{bpm-vpmeas_vlf160} }
\subfigure[]{ \includegraphics[width=0.45\textwidth]{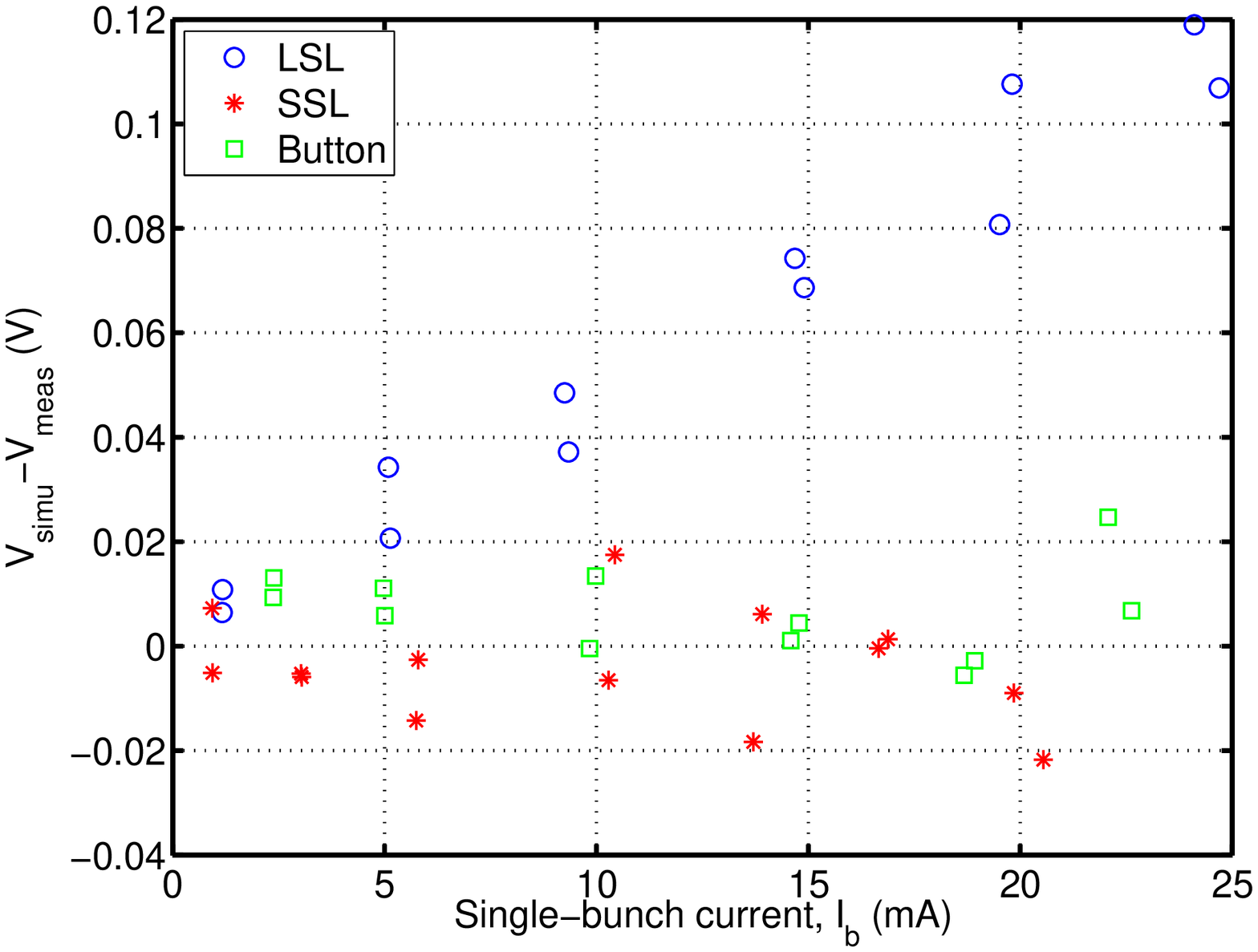}
\label{bpm-vperr_vlf160} }
\subfigure[]{ \includegraphics[width=0.432\textwidth]{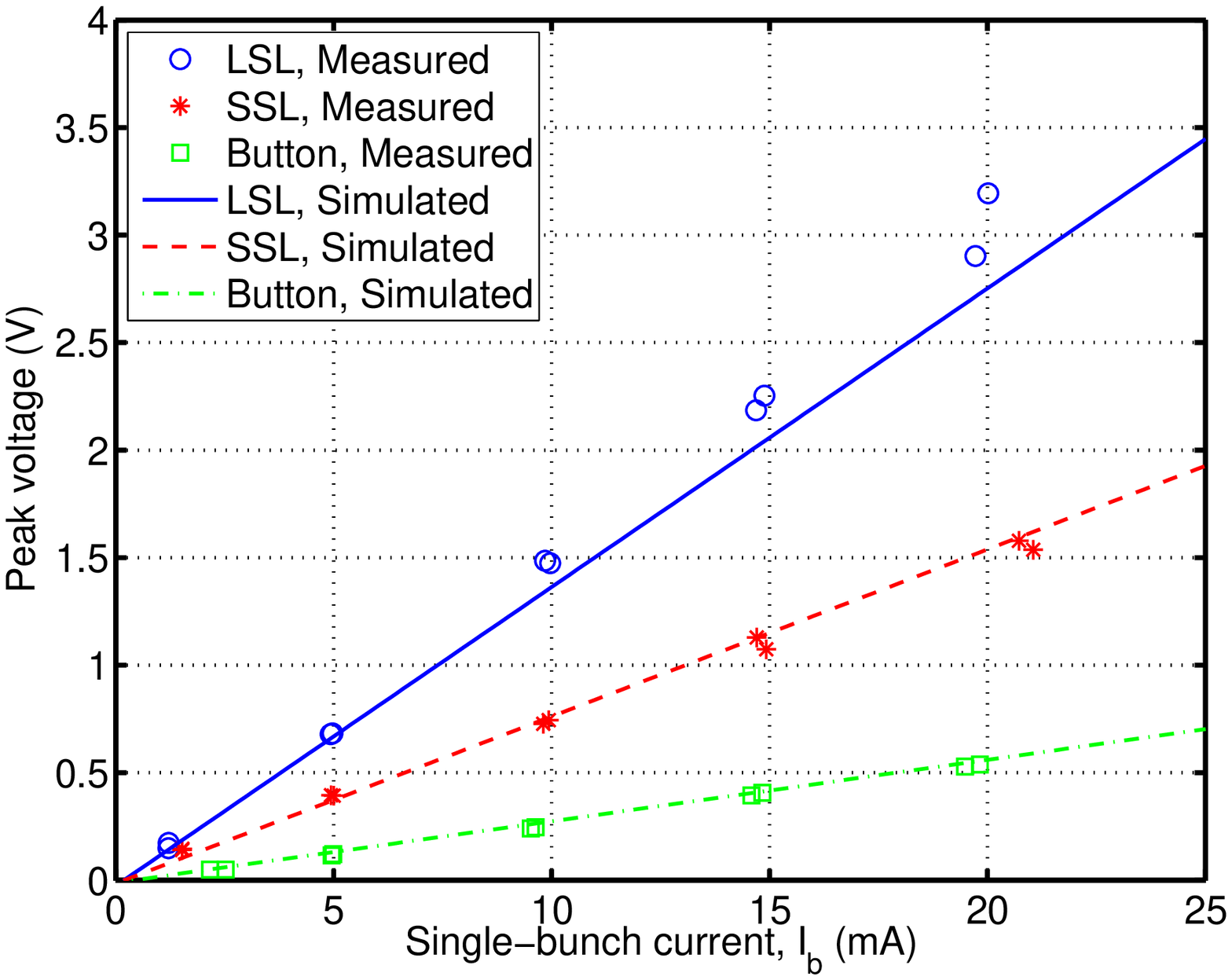}
\label{bpm-vpmeas_vlf320} }
\subfigure[]{ \includegraphics[width=0.45\textwidth]{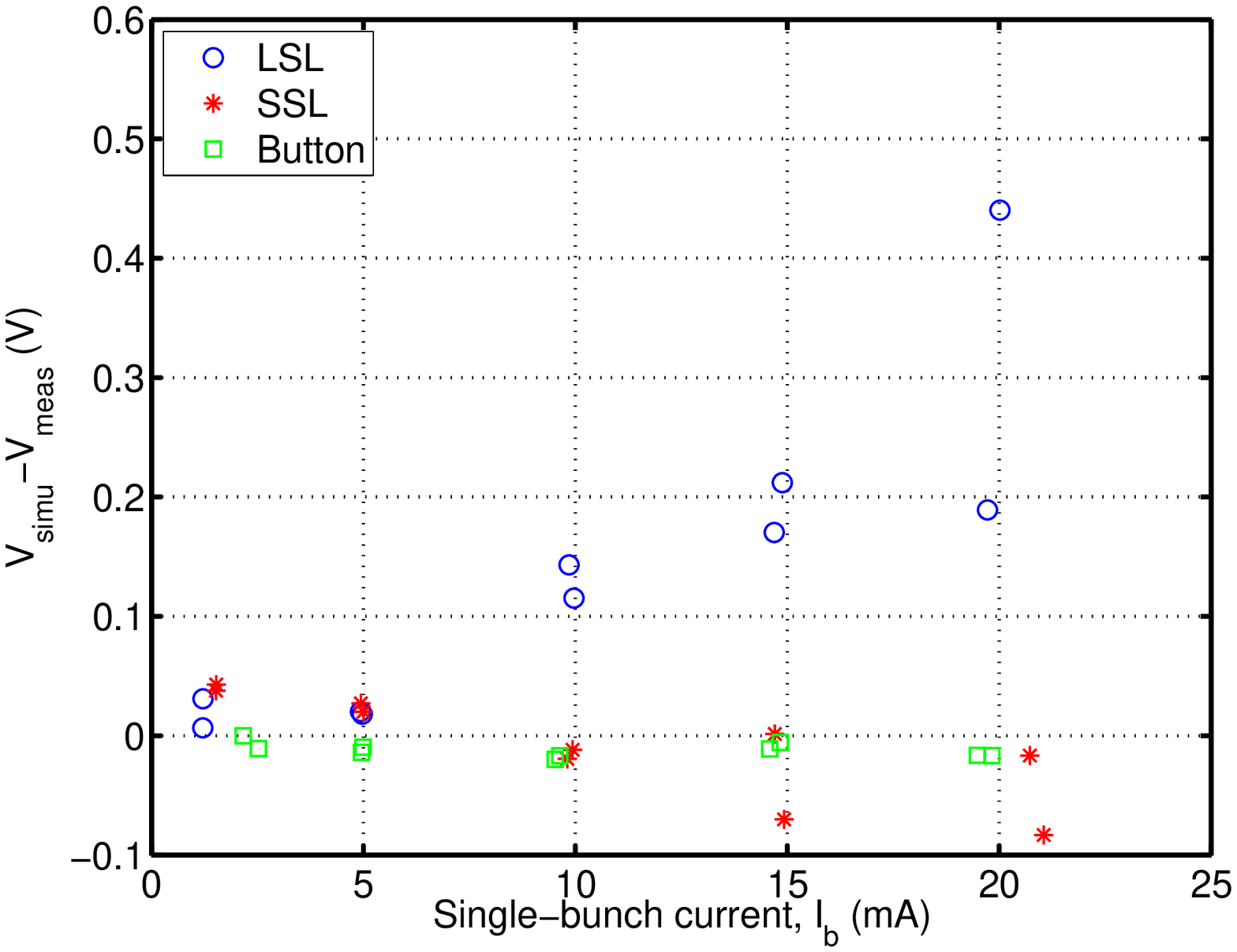}
\label{bpm-vperr_vlf320} }
\caption{(color online) The beam based test results of the signal conditioning vs the
simulated values.
(a) Peak voltages of the pickup signal conditioned using VLF-160 LPFs;
(b) peak voltage differences of the beam based measurement and simulated values for VLF-160 LPFs;
(c) peak voltages of the pickup signal conditioned using VLF-320 LPFs; and
(d) peak voltage differences of the beam based measurement and simulated values for VLF-320 LPFs.
}
\label{bpm-vp_meas}
\end{figure*}
is tested using the electron beam in the storage ring. A 60-feet long RG232-U
cable is used to bring signals from BPM pickups to the TD7404 Tektronics
digital oscilloscope. A VLF-160 or VLF-320 LPF is connected on
the oscilloscope end of the cable.
This arrangement can reduce the impact
of the reflected waves on the electron beam.
The measured peak voltages of the conditioned pickup signals using either VLF-160 or VLF-320
LPFs are shown in Fig.~\ref{bpm-vpmeas_vlf160} and ~\ref{bpm-vpmeas_vlf320}, respectively.
Figure~\ref{bpm-vperr_vlf160} and~\ref{bpm-vperr_vlf320} show the differences between the measured
peak voltages and simulated values.
For button and SSL pickup signals conditioned using either VLF-160 LPFs or VLF-320 LPFs,
the absolute differences of the measured and calculated peak voltages are within $\pm 0.10$ V
when the single-bunch current is increased from 2 mA to 25 mA.
The absolute peak voltage differences of an LSL pickup are less than 0.12 V when conditioned using
a VLF-16 LPF, and less than 0.45 V when conditioned using a VLF-320 LPF,
in the single-bunch current range from 2 mA to 25 mA.
These results indicate that the BPM electronics is expected to work
properly with a single-bunch current up to 100 mA for all three types of pickups
when conditioned using VLF-160 LPFs,
and for button and SSL pickups when conditioned VLF-320 LPFs.
The beam based measurements are consistent with the simulated results.

\section{Beam based BPM measurement}

The ultimate goal for the BPM pickup signal conditioning is to increase
the BPM's dynamic range and keep the electron beam orbit
stable for FEL and HIGS operations.
This section reports the studies on the BPMs' performance in various operation conditions.

A set of four VLF-160 or VLF-320 LPFs are used for each BPM.
These four LPFs are matched based upon careful measurements of their insertion losses in
the BPM's operation range using a network analyzer. VLF-320 LPFs are
only used for two button BPMs (S01 and S04) in the south straight section.
For all other BPMs, VLF-160 LPFs are employed for the signal conditioning.
The LPFs are installed at the BPM electronics module end to minimize the
impact of the reflected signal on the electron beam.

\subsection{Thermal effect induced orbit drifting}
The measured electron beam orbit in the storage ring would
drift with time if the slow global orbit feedback is turned off.
This orbit drifting effect makes it difficult to study the dynamic range of the BPM system.
To better understand the orbit drifting effect,
a series of measurements are performed.
\begin{figure}[tb]
\centering \subfigure[]{ \includegraphics[clip,width=0.45\textwidth]{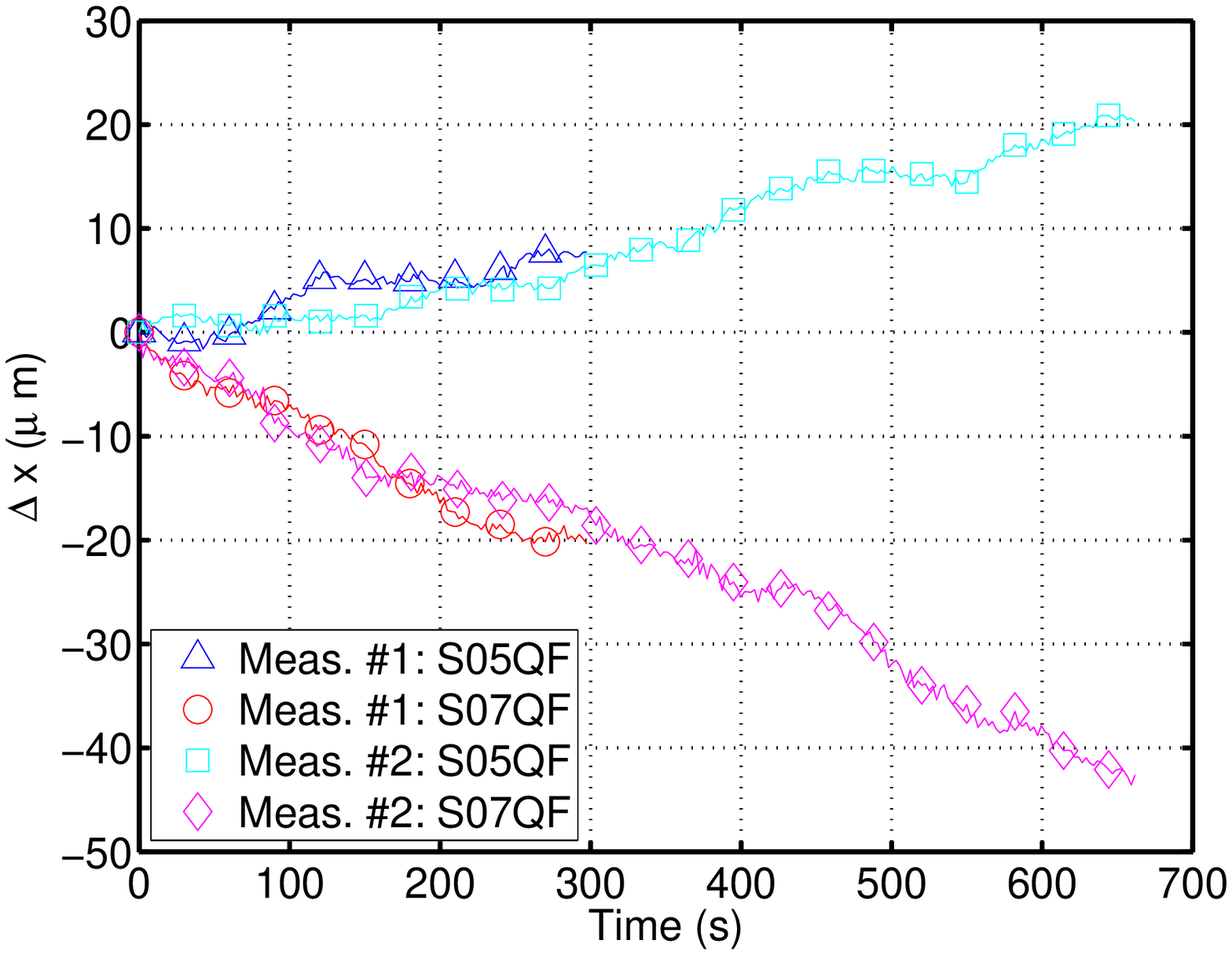}
\label{drift1} }
\subfigure[]{ \includegraphics[clip,width=0.45\textwidth]{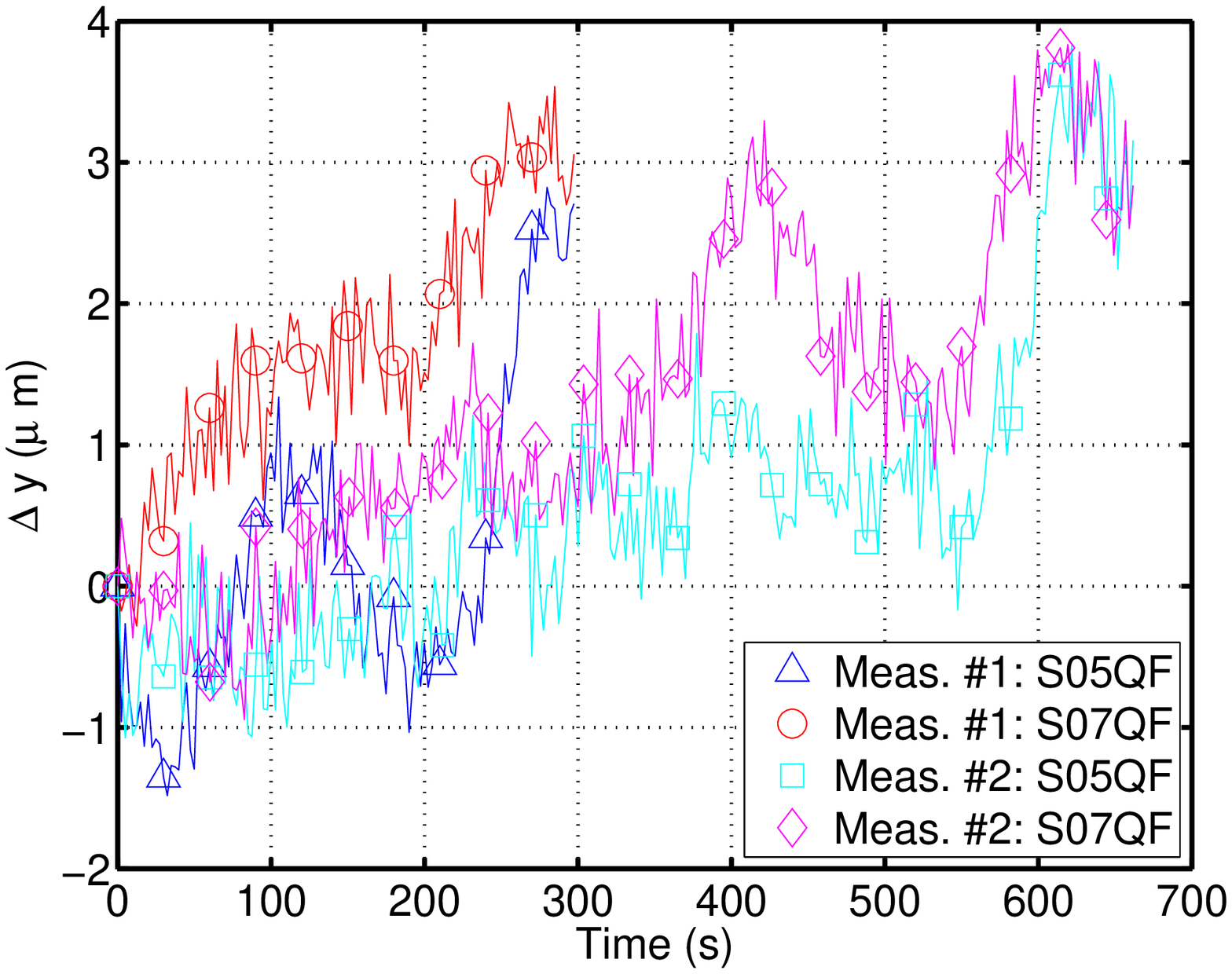}
\label{drif2} }
\caption{(color online) The orbit drifting around the electron-photon collision point.
The global orbit feedback is turned off during the measurements.
In measurement \#1, the electron beam decays from 9.0 mA to 8.1 mA in 5 minutes.
In measurement \#2, the electron beam decays from 4.8 mA to 4.0 mA in 11 minutes.
 (a) The horizontal orbit; (b) the vertical orbit.}
\label{orb_drift}
\end{figure}

\begin{table}[hbt]
\centering
\caption{Maximum orbit drift measured  as a function of time using BPMs with different types
of pickups.
In measurement \#1, the electron beam decays from 9.0 mA to 8.1 mA in 5 minutes.
In measurement \#2, the electron beam decays from 4.8 mA to 4.0 mA in 11 minutes.}
\label{tbl-orb_drift}
\begin{tabular}{lcccc}
\toprule
\multirow{2}{*}{} & \multicolumn{2}{c}{Meas. \#1 ($\mu$m)} & \multicolumn{2}{c}{Meas. \#2 ($\mu$m)} \\
\cline{2-5}
& $\Delta$ x &  $\Delta$ y &   $\Delta$ x &  $\Delta$ y\\
\hline
LSL   & 10 &5 &21&9 \\
SSL   & 21 &5 &44&10 \\
Button& 12 &11&26&24 \\
Collision point   & 21 &5&44&6 \\
\bottomrule
\end{tabular}
\end{table}

Many factors can contribute to the orbit drifting.
One important effect is temperature related.
For example, the environment temperature variation can affect the BPM readings and
cause magnets movement which leads to changes of the electron beam closed orbit~\cite{HUBERT}.
Table~\ref{tbl-orb_drift} shows the maximum orbit drifts measured using different types
BPMs in two consecutive measurements.
The global orbit feedback is turned off during the measurements.
In measurement \#1, a single-bunch current decays from 9.0 mA to 8.1 mA in 5 minutes.
In measurement \#2, the bunch current decays from 4.8 mA to 4.0 mA in 11 minutes.
The orbit drifts around the electron-photon collision point measured by
S05QF and S07QF BPMs in these two measurements are shown in Fig.~\ref{orb_drift}.
The results show that the orbit drifts as a function of time,
and varies tens of microns in about 10 minutes.
More tests indicate that the drifting
trends and amplitudes in different time windows are different.

\subsection{Dependency on single-bunch current }

\begin{table}[b]
\centering
\caption{Maximum orbit change measured as a function of the single-bunch current using BPMs with
different types of pickups.
In measurement \#1, the electron beam decays from 28.6 mA to 4.9 mA in 4.3 minutes.
The electron beam decays from 31.8 mA to 6.4 mA in 3.5 minutes in measurement \#2.}
\label{tbl:Ib_orb}
\begin{tabular}{lcccc}
\toprule
\multirow{2}{*}{} & \multicolumn{2}{c}{Meas. \#1  ($\mu$m)} & \multicolumn{2}{c}{Meas. \#2  ($\mu$m)} \\
\cline{2-5}
& $\Delta$ x &  $\Delta$ y&   $\Delta$ x &  $\Delta$ y \\
\hline
ARC   & 116 &14 &127&15 \\
NSS   & 15 &15 &11&15\\
SSS& 33 &25&25&15 \\
Collision point   & 24 &12&17&12 \\
\bottomrule
\end{tabular}
\end{table}

\begin{figure}[!htb]
\centering \subfigure[]{ \includegraphics[clip,width=0.45\textwidth]{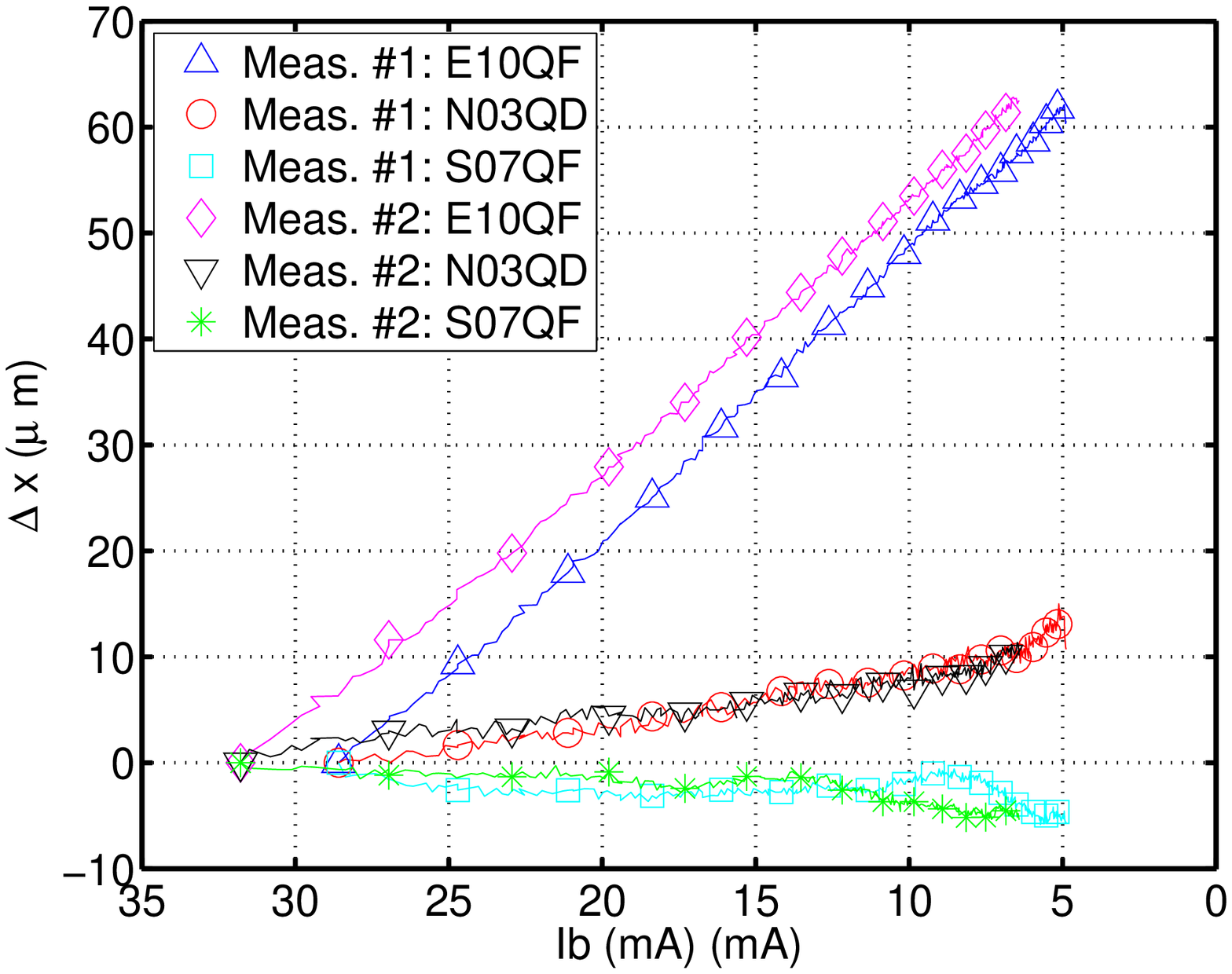}
\label{col-angle1} }
\subfigure[]{ \includegraphics[clip,width=0.45\textwidth]{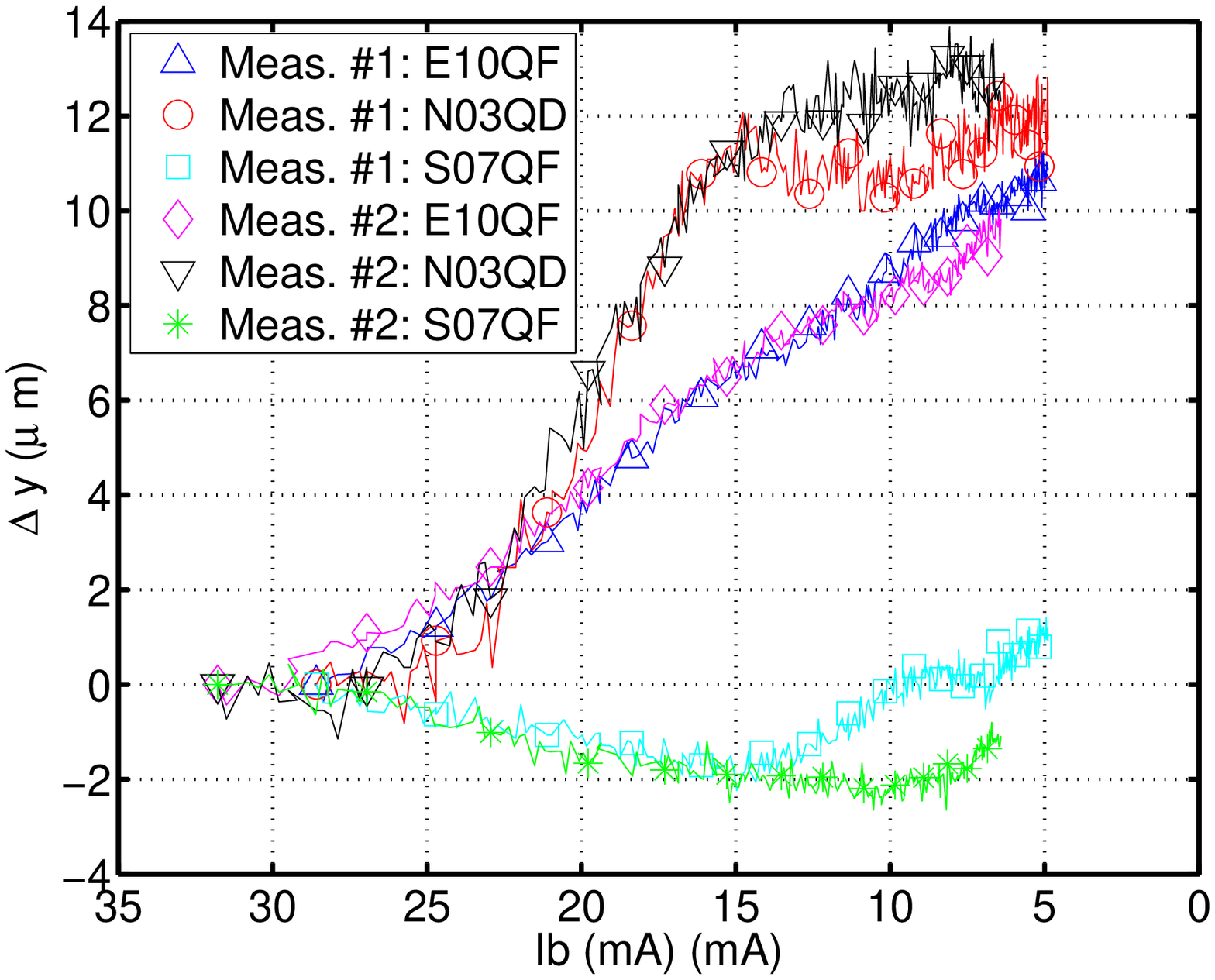}
\label{col-angle2} }
\caption{(color online) Dependency of the BPM measurement on the single-bunch current.
       In measurement \#1, a single-bunch current decays from 28.6 mA to 4.9 mA in 4.3 minutes.
       In measurement \#2, a single-bunch current decays from 31.8 mA to 6.4 mA in 3.5 minutes.
       BPM E10QF, N03QD and S07QF are located in the east arc, NSS and SSS, respectively.
}
\label{fig:Ib_orb}
\end{figure}
According to the test results in the previous section, it is difficult to measure the actual
dependency of the BPM readings on the single-bunch current because of the orbit drifting
due to thermal effects.
In order to reduce the orbit drifting effect,
the test duration should be as short as possible.
This is realized by lowering the RF voltage to about 60 kV in our tests to reduce the beam lifetime,
and thus to shorten the test duration.
Table~\ref{tbl:Ib_orb} lists the results of two measurements for the three types of
BPMs and the BPMs around the electron-photon collision point.
Figure~\ref{fig:Ib_orb} shows the orbit changes of three BPMs located in different sections
as a function of the electron beam current.
In the first measurement, a 28.6 mA single-bunch
beam is injected into the storage ring.
Then,  the orbit is recorded as the beam current decays to 4.9 mA.
The duration for this test is 4.3 minutes.
In the second measurement, the single-bunch beam decays from 31.8 mA to 6.4 mA,
and the orbit is monitored for 3.5 minutes.
In these measurements,
the horizontal readings of the arc BPMs significantly depend on the single-bunch current.
This effect is attributed to the large horizontal orbit offset  ($\sim$2.5 mm) in the arc BPMs
necessary for using combined quad-sextupole in the arc.
The measured vertical orbit changes in the arcs have similar amplitudes as the thermal drifts
in this period of time.
The orbit deviation with the single-bunch current in straight sections,
for both horizontal and vertical directions,
also have similar range as the thermal drifting in the same time duration.
These results indicate that the dependency of the BPM readings on the single-bunch current
is small in the straight sections,
and thus has little impact on the light source operation since the FELs
and electron-photon collision point are all located in the SSS.

\subsection{Dependency on multi-bunch current}

\begin{figure}[!htb]
\centering
\subfigure[]{ \includegraphics[clip,width=0.45\textwidth]{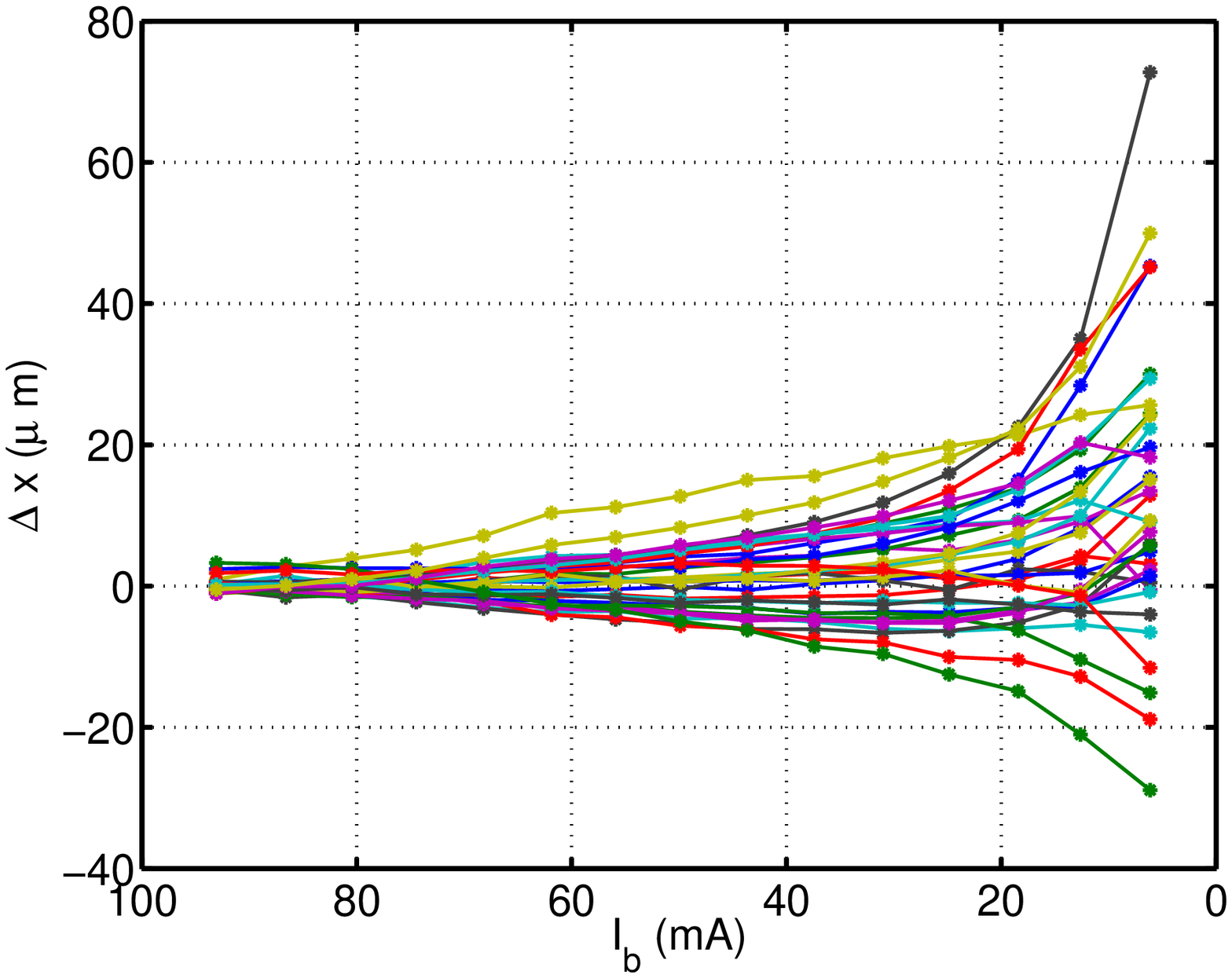}
\label{orb-currA1} }
\subfigure[]{ \includegraphics[clip,width=0.45\textwidth]{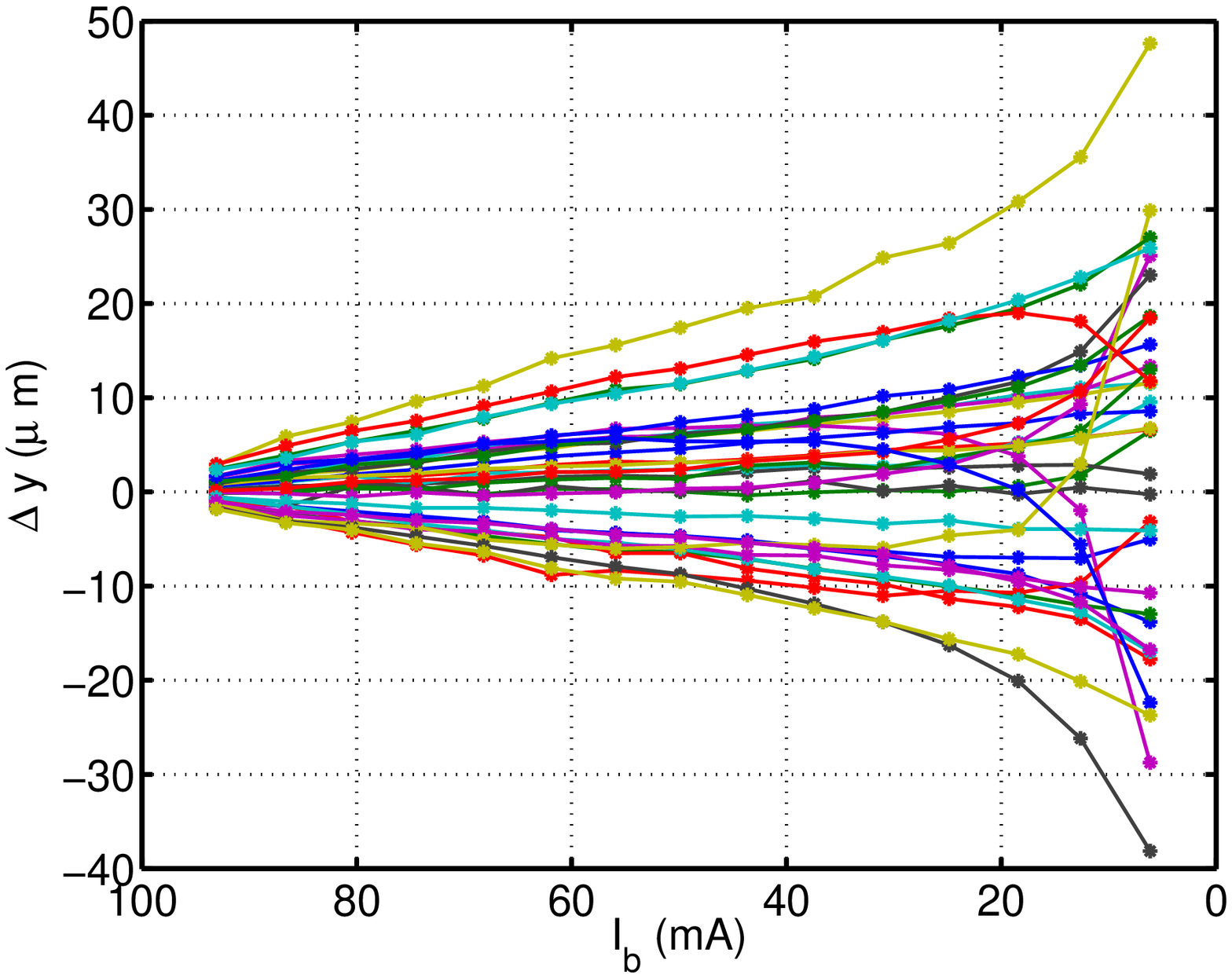}
\label{orb-currA2} }
\caption{(color online) The BPM readings measured as a function of the beam current.
A 100 mA electron beam with 16 evenly distributed bunches are
injected in the storage ring.
Then the stored bunches are killed one-by-one using the TFB system.
Every line represents the BPM readings of a particular BPM.
(a) Horizontal BPM readings; (b) vertical BPM readings.
}
\label{orb-currA}
\end{figure}

\begin{figure}[!htb]
\centering
\subfigure[]{ \includegraphics[clip,width=0.45\textwidth]{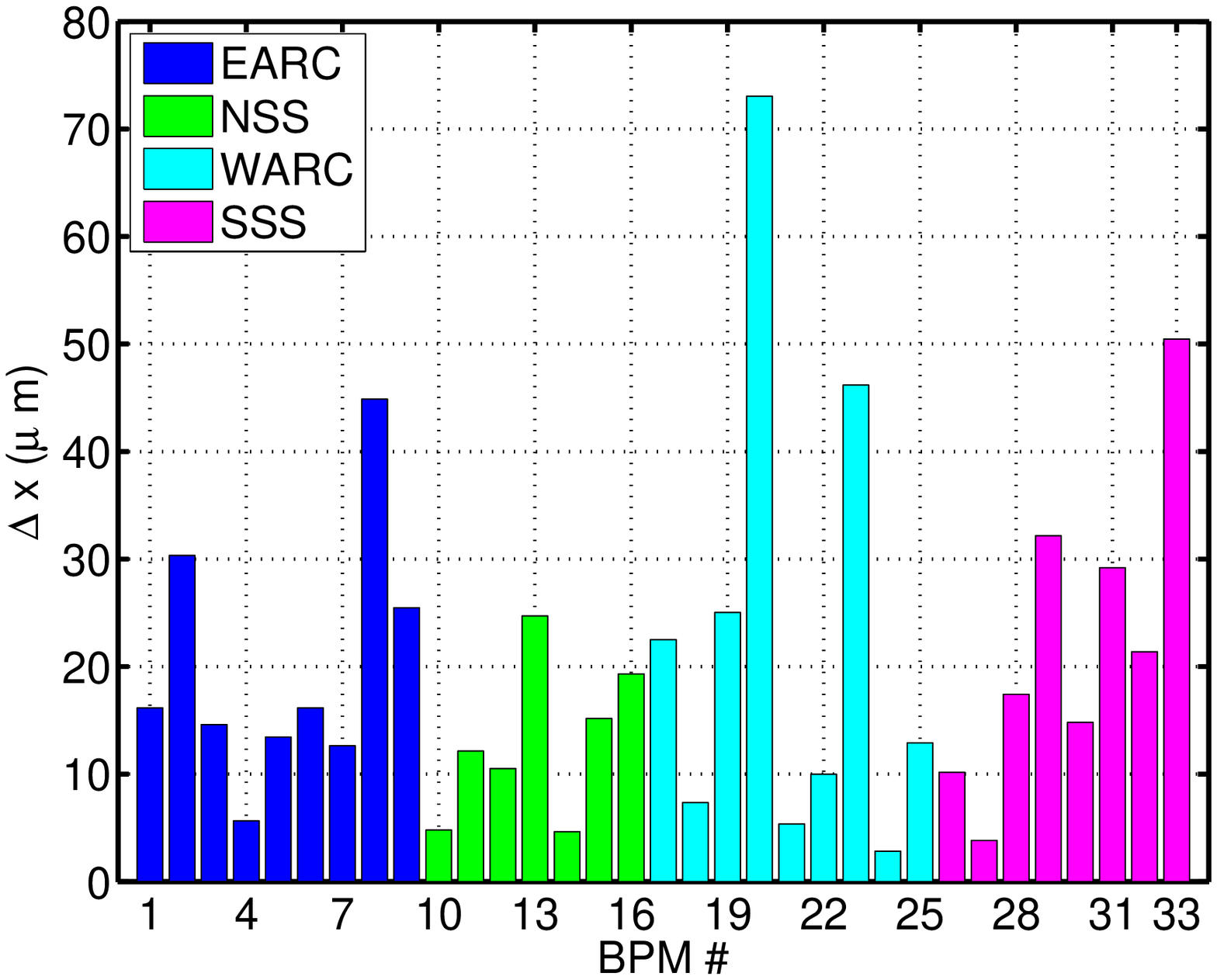}
\label{orb-currB1} }
\subfigure[]{ \includegraphics[clip,width=0.45\textwidth]{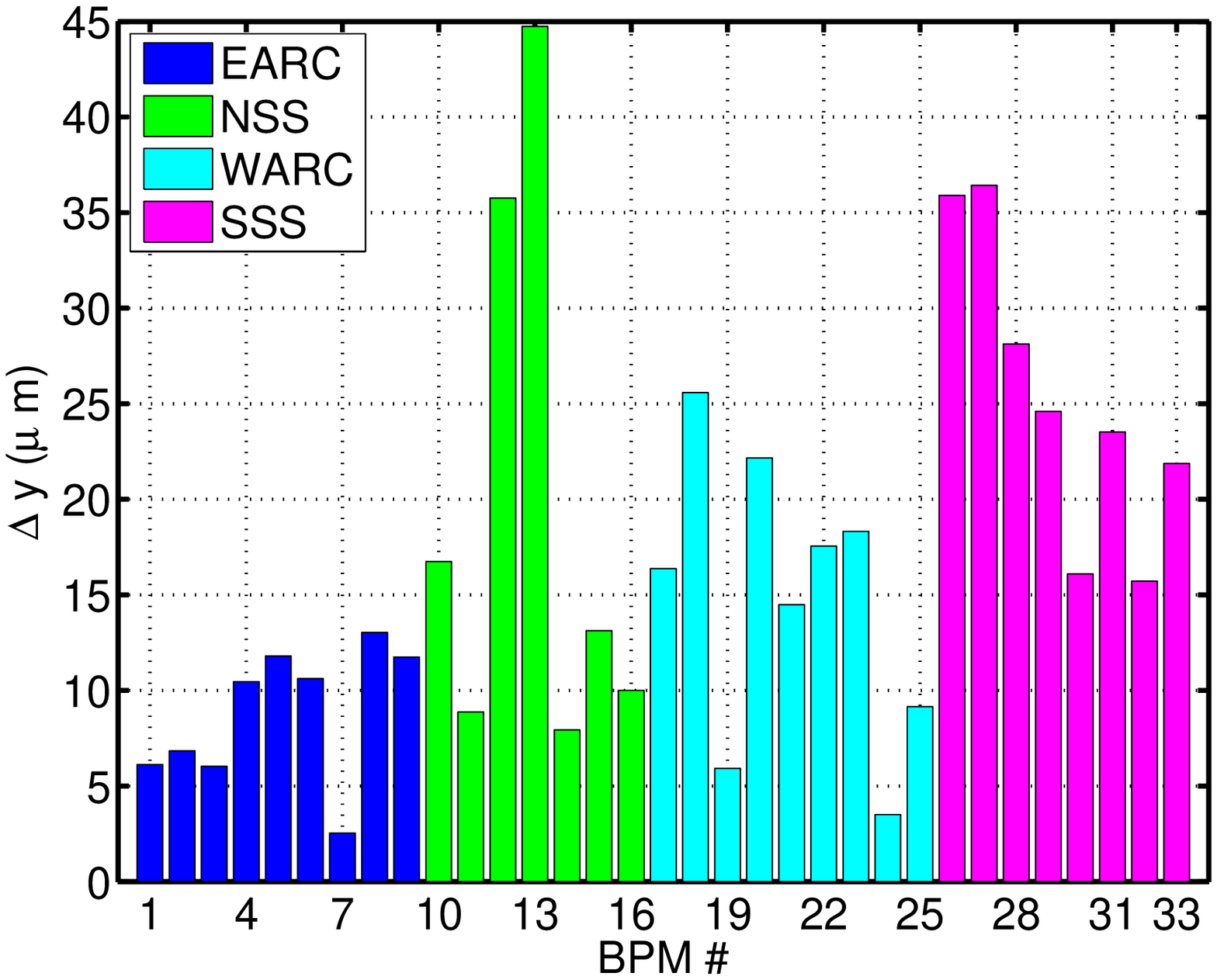}
\label{orb-currB2} }
\caption{(color online) The maximum changes of the BPM readings as the beam
current is reduced from 100 mA to 6 mA in the same test as shown in Fig.~\ref{orb-currA}. }
\label{orb-currB}
\end{figure}
The Duke storage ring is planned to be operated in multi-bunch modes to achieve a higher
$\gamma$-ray flux.
To learn about the behavior of the BPMs at different multi-bunch current and
in different bunch patterns,
a series of studies are performed using
a recently developed beam cleaning method based upon the bunch-by-bunch transverse feedback
(TFB) system~\cite{WWZ2012}.
In each test, an electron beam with 16 evenly distributed bunches is
injected into the storage ring.
Then the stored bunches are killed one-by-one using the TFB system.
The electron beam orbit and current are recorded before and after killing
each bunch.
In this test, a 100 mA, 16-bunch electron beam is evenly injected into the storage ring.
The duration of this test is about 70 seconds,
and the thermal effect induced orbit drift is small in this short period of time.
Figure~\ref{orb-currA} shows the measured orbit changes as a function of the beam current
in one of the tests,
and Figure~\ref{orb-currB} shows the maximum orbit changes of the same measurement.
The results indicate when the beam current changes from 100 mA to 80 mA,
the maximum orbit variations are less than 4 $\mu$m and  8 $\mu$m
in horizontal and vertical directions, respectively.
At the Compton collision point in the SSS,
the typical horizontal RMS beam size is about 370 $\mu$m.
With estimated 5\% emittance coupling,
the typical vertical RMS beam size at the same location is about 80 $\mu$m.
These BPM reading changes are less than 10\% of the beam size at the collision point.
As the HIGS $\gamma$-ray production is run typically with top-off injection to keep electron
beam current steady,
the stability of BPM readings is more than adequate for the $\gamma$-ray
production in the top-off mode of operation.
The result also indicates that
most of the measured orbit changes are within $\pm$20 $\mu$m
in both horizontal and vertical directions when the total beam current is between 100 mA and 20 mA,
and the maximum BPM reading changes are less than 50 $\mu$m in the current range from 100 to 6 mA
for the BPMs in the straight sections.

\section{Summary}

The Duke storage ring has a number of operation modes with the single-bunch
current varies from about 1 mA to 100 mA.
To obtain precise orbit measurements in a wide range of the single-bunch current,
a method for conditioning the BPM pickup signal is developed to enlarge
the dynamic range of the BPMs.
In this method, low pass filters are employed to effectively
reduce the peak voltage of the pickup signals,
and make the BPM system  capable of providing reasonable orbit measurements
with a single-bunch current of 1 to 100 mA.

A number of tests under different operation conditions are performed
to check the performance of the BPM system with signal conditioning.
The dependency of the straight section BPM readings on the single-bunch current is small,
and cannot be precisely measured due to thermal effects.
The arc BPMs show larger orbit reading variations as the single-bunch current is changed.
But this has little impact on most operations of the HIGS and FELs.
The dependency of the BPM readings on the multi-bunch current and bunch pattern is also tested.
The results indicate that multi-bunch current and bunch pattern do not significantly
impact the BPM readings as long as the bunches are evenly filled
and the total current is kept in a reasonable range (20\%--30\%).
\\

\acknowledgments{The authors would like to thank all colleagues at Duke FEL laboratory,
TUNL, for their technical and operational supports.}

\end{document}